\documentstyle[aps]{revtex}
\begin{document}
\draft
\title{
Magnetic and electronic structures of superconducting RuSr$_2$GdCu$_2$O$_8$
}
\author{K.~Nakamura, K.~T.~Park and A.~J.~Freeman}
\address{
Department of Physics and Astronomy, Northwestern University, Evanston, IL60208}
\author{J.~D.~Jorgensen}
\address{
Materials Science Division, Argonne National Laboratory, Argonne, IL 60439}

\date{\today}
\maketitle
\begin{abstract}
The coexistence of ferromagnetism and superconductivity in 
RuSr$_2$GdCu$_2$O$_8$ was reported both from experiments (by Tallon et. al.) 
and first-principles calculations (by Pickett et. al.).
Here we report that
our first-principles full-potential linearized augmented  
plane wave (FLAPW) calculations, 
employing the precise crystal structure with 
structural distortions (i.e., RuO$_6$ rotations) determined by neutron diffraction, 
demonstrate that antiferromagnetic ordering of the Ru moments is 
energetically favored over the previously proposed ferromagnetic ordering. 
Our results are consistent with 
recently performed magnetic 
neutron diffraction experiments (Lynn et. al).
Ru $t_{2g}$ states, which are responsible for the magnetism,
have only a very small interaction with Cu $e_g$ states,
which results in a small exchange splitting of these states.
The Fermi surface, characterized by strongly hybridized $dp\sigma$ orbitals, 
has nesting features 
similar to those in the two-dimensional high $T_c$ cuprate superconductors. 
\end{abstract}

\vspace{1.0cm}
%
%
\section{Introduction}
The reported discovery of the coexistence of ferromagnetism and superconductivity 
by Tallon et. al. \cite{99Tal,99Ber} 
has attracted a great deal of interest, 
since ferromagnetism breaks the degeneracy of spin-up and spin-down partners 
of Cooper pairs.\cite{57Gin}
The experiments demonstrated that RuSr$_2$GdCu$_2$O$_8$ exhibits 
ferromagnetic ordering of Ru moments 
below a Curie temperature, $T_C \approx $ 133~K, 
and becomes superconducting at a lower temperature, $T_c$ up to 40~K. 
From a theoretical point of view, Pickett et. al.\cite{99Pic} calculated 
the electronic structure of RuSr$_2$GdCu$_2$O$_8$
and discussed 
how the superconductivity, 
which may behave as a Fulde-Ferrell-Larkin-Ovchinniko 
(FFLO)
type superconductor,
can coexist with the ferromagnetism.
However, 
the X-ray determined crystal structure used in their calculations
was different from that recently determined by neutron diffraction\cite{00Chm}
- and so they did not consider the structural distortions (i.e., oxygen rotations) 
in the RuO$_6$ octahedra.

The crystal structure of RuSr$_2$GdCu$_2$O$_8$ 
with a P4/mbm tetragonal space group, 
determined by the neutron experiment,\cite{00Chm} 
is shown in Fig.~\ref{fig1}~(a). 
This is a structure similar to that of YBa$_2$Cu$_3$O$_7$, 
where Y, Ba and Cu (chain atom) are replaced by Gd, Sr and Ru, respectively. 
Ru lies at a six-coordinated position in the octahedron composed 
of six neighboring oxygens (four O$_{Ru}$ and two O$_{apical}$), 
while Cu lies in a five-coordinated position (four O$_{Cu}$ and one O$_{apical}$). 
Since the interatomic distance of Cu-O$_{apical}$ (2.190~\AA) 
is much larger than that of Ru-O$_{apical}$ (1.912~\AA), 
the Cu has a rather weak interaction with O$_{apical}$. 
Thus, this may yield a two-dimensional (2D) electronic structure 
of the Cu-O$_{Cu}$ layer.
The RuO$_6$ octahedra are rotated by about 14$^\circ$ 
around the c-axis to fill the space more efficiently, 
as shown in Fig.~\ref{fig1}~(b).
These rotations may lead to 
a significant reduction in the $d$-electron bandwidth 
due to a deviation of the angle \angle Ru-O$_{Ru}$-Ru from 180$^\circ$,
and so the magnetism of Ru may be strongly perturbed by the structural distortion. 
Hence, it is necessary to again 
investigate the stability of the magnetic structure in RuSr$_2$GdCu$_2$O$_8$
by first-principles calculations with the precise crystal structure parameters.

With the experimentally determined structure,\cite{00Chm}
our highly precise first-principles 
full-potential linearized augmented plane wave(FLAPW) calculations\cite{81Wim}
on three different magnetic orderings
demonstrate\cite{00Nak} that 
antiferromagnetic ordering of the Ru moments is 
energetically favorable over the previously proposed ferromagnetic ordering.
Furthermore, 
in agreement with our result,
recent neutron diffraction experiments by Lynn et. al.\cite{00Lyn} 
observed antiferromagnetic Ru ordering.
In the present paper, we report the first-principles results 
on RuSr$_2$GdCu$_2$O$_8$, and discuss the magnetic and electronic structures
and their implications for the coexistence of superconductivity.

\section{Model and calculation method}
In our FLAPW calculations, we employed the experimentally determined 
room temperature crystal structure.\cite{00Chm}
For the magnetic structures,
we assumed three kinds of magnetic orderings 
for Ru and Gd moments as depicted in Fig.2:
(a) a FM structure with ferromagnetic orderings for both Ru and Gd moments;
(b) an AFM-I structure with a C-type antiferromagnetic ordering of Ru moments 
and a ferromagnetic ordering of Gd moments
(more precisely, a ferrimagnetic structure);
and (c) an AFM-II structure with C-type antiferromagnetic orderings 
for both Ru and Gd moments.
(Note that the symmetry of both the AFM-I and AFM-II structures makes 
two inequivalent sublattice sites for Ru, O$_{apical}$ and Cu.)
While the recent experiment\cite{00Lyn} revealed 
antiferromagnetic alignments for both Ru and Gd moments with G-type orderings, 
in which nearest neighbors of Ru (Gd) moments 
along all three crystallographic axes are coupled antiferromagnetically,
we employed the C-type antiferromagnetic ordering
in order to reduce the large computational effort this would entail.
(The G-type ordering requires doubling the unit cell of the C-type ordering.)
This is justified because the moment alignment along the c-axis may be less important 
than that along the a-axis
since the distance between the neighboring Ru (Gd) atoms in the c-axis (11.56~\AA)
is significantly greater than that in the a-axis (3.84~\AA).
Thus, the assumption of C-type magnetic structure would not  
alter the results or the insight obtained.

Calculations were performed based on the local spin density approximation (LSDA)
with the Hedin-Lundquist exchange correlation\cite{71Hed},
in which the core states are treated 
fully relativistically and the valence states are treated semi-relativistically. 
Although the effects of electronic correlation in a strongly correlated system 
may be taken into account within a scheme such as LDA+U,
RuSr$_2$GdCu$_2$O$_8$ shows metallic character even in the RuO layer, 
as will be presented in Sec. IV, which causes that effect to be weak.
Hence, while our results are restricted within LDA, 
they may be sufficient to discuss qualitative conclusions.
The LAPW basis functions were used with a cut-off, $|k+G|<3.6$ a.~u., 
corresponding to about 1800 plane waves.
Muffin-tin (MT) sphere radii in a.~u. were chosen as: 
2.4 (Ru), 2.5 (Sr), 2.5 (Gd), 2.2 (Cu) and 1.2 (O).
Inside the spheres, the angular momentum expansion was truncated at $l=8$ 
for wave functions, charge density and potential.
Integrations were performed over a first Brilliouin zone 
using 126 special $k$-points, 
corresponding to 16, 16 and 32 $k$-points 
in the irreducible Brilliouin zone 
for the FM, AFM-I and AFM-II structures, respectively.

\section{Results}
\subsection{Magnetic Moments}
The calculated magnetic moment of each atom inside the MT sphere 
for the FM, AFM-I and AFM-II structures
is given in 
Table~\ref{table1}.
The moment of Gd is close to 7~$\mu_B$ - as expected.
In the AFM-I structure, 
the magnitude of the Ru, O$_{apical}$ and Cu magnetic moments 
with opposite spin directions is slightly different 
since the ferromagnetically ordered Gd moments 
break the symmetry of the two sublattice sites. 
Among all three structures, 
the Ru magnetic moments are almost of the same magnitude, 
about 1.55~$\mu_B$. 
Note that since the magnitude depends on the MT sphere radius,
the rather large MT sphere radius (2.4~a.u.) we used gives rather large values.
However, the experimental value (1.18~$\mu_B$) of the Ru magnetic moment 
determined by neutron diffraction\cite{00Lyn}
is smaller than our results. 
This may be attributed to possible canted antiferromagnetic ordering
of the Ru magnetic moments,
for which further investigations are necessary.
The induced magnetic moments of O$_{apical}$
and O$_{Ru}$ are very sensitive to the Ru magnetic structures,
while the Cu and O$_{Cu}$ moments are negligibly small.

\subsection{Total energy}
Total energies for all three FM, AFM-I and AFM-II structures are also given in 
Table~\ref{table1}.
From a comparison between the FM and AFM-I structures, 
we found that an antiferomagnetic ordering of Ru moments 
is favored over the ferromagnetic ordering. 
The total energy difference, 22.5~meV/Ru-atom, between both FM and AFM-I
corresponds to the Ne\'{e}l temperature ($T_N$) of 73~K, 
crudely estimated by mean field theory with $S=3/2$.
Although the calculated $T_N$ is lower than 
the experimentally determined $T_N$,  136~K,\cite{00Lyn}
our prediction for the magnetic structure is consistent with 
the experiment by Lynn et. al.\cite{00Lyn}
Further, we found that an antiferromagnetic ordering of 
the Gd moments (AFM-II) reduces the total energy,
but only by 2.3~meV/Gd-atom, from that of the AFM-I structure, 
corresponding $T_N\approx 7$~K (assumed $S=7/2$),
which is approximately in agreement 
with the observed low Ne\'{e}l temperature, 2.5~K,\cite{00Lyn} 
for the Gd moment alignment.

\section{Discussion}
\subsection{Density of states}
Figure~\ref{fig3} shows
the total density of states (DOS) and partial DOS of each atom 
for the FM, AFM-I and AFM-II structures. 
The peaks at 2~eV above the Fermi level ($E_F$) and 
3~eV below $E_F$ in the total DOS 
for all three structures correspond to Gd 4$f$ states which are strongly localized,
and their states do not significantly affect the electronic structure around $E_F$.
An exchange splitting in the total DOS in the AFM-I structure is 
due to the ferromagnetism of Gd,
but the difference between majority and minority spin DOS is very small 
except in the region of the localized Gd 4$f$ states.
Note that the total DOS for all three structures shows metallic character.

From the partial DOS in Fig.~\ref{fig3},
we can see that the magnetism of Ru is dominated by antibonding $t_{2g}$ states,
and the major O$_{Ru}$ and O$_{apical}$ peaks correlate strongly 
with those of Ru.
In contrast, the partial DOS for Cu $d$ and O$_{Cu}$ $p$ states, 
which are strongly hybridized, 
are insensitive to the magnetism of Ru. 
The exchange splitting of the Ru $d$ states 
for all three structures is about 1~eV, 
while the induced exchange splitting in the Cu $d$ states is
smaller by two orders of magnitude - 0.01 to zero eV.
Our results for the exchange splitting and magnetic moments of the Cu $d$ states
is smaller, even in the FM case, than 
those in the undistorted FM case reported by Weht et. al.\cite{99Weh} 
This may be due to the differernt structure parameters 
employed in the two calculations. 
The fact that the Ru-Cu distance, 4.10 \AA,
in our calculations is larger than in their case, 3.58 \AA, 
results in a rather weak magnetic interaction
and leads to smaller exchange splittings and magnetic moments in the Cu $d$ states. 
Futher, the relative atomic position of O$_{apical}$ 
between the Ru and Cu may be sensitive to the magnetism.\cite{99Weh} 
However, the exchange splitting and the magnetic moments in the Cu $d$ are quite small.
The small exchange coupling is due to the unique electronic structure of 
the layered Ru $t_{2g}$ and Cu $e_g$ states 
separated by $p$ orbitals of O$_{apical}$, 
as discussed by Pickett et. al.,\cite{99Pic} 
and is roughly valid regardless of the magnetic ordering of Ru moments:
Ru $t_{2g}$ states couple only to 
the $p_x$ and $p_y$ orbitals of O$_{apical}$, 
but do not couple to the Cu-O $d_{x^2-y^2}$ states.

\subsection{Magnetism of Ru}
We focus here on the magnetism of Ru.
Figure~\ref{fig4} shows 
the projected DOS in real space for Ru $d$, O$_{Ru}$ $p$ and O$_{apical}$ $p$ 
states for the FM, AFM-I and AFM-II structures. 
The $x$ and $z$-axes for the Ru (O$_{Ru}$) DOS
are chosen as directions to the neighboring O$_{Ru}$ (Ru) site 
and c-axis, respectively.
Compared to the DOS of the FM structure presented by Weht et. al.,\cite{99Weh}
which did not include a structural distortion due to RuO$_6$ rotations,
our calculated majority $d_{xy}$ states (cf., Fig.~\ref{fig4} (a)) are 
more localized below $E_F$,
due to deviation of the angle \angle Ru-O$_{Ru}$-Ru from 180$^\circ$ 
and the elongation of Ru-O$_{Ru}$ bonds by the RuO$_6$ rotations.
This results in the $d_{xy}$ states being almost fully occupied,
and in a depletion of the majority spin hybridization channel at $E_F$.
In the AFM-I and AFM-II structures,
the majority $d_{xy}$ and $d_{xz(yz)}$ states on a Ru site
can hybridize with the minority spin states on neighboring Ru sites 
through O$_{Ru}$ $p_y$ and O$_{Ru}$ $p_z$ orbitals, respectively, 
by the superexchange mechanism.
The $d_{xz(yz)}$ states also couple strongly to the O$_{apical}$ $p_x(y)$ orbitals.
The charge configuration of Ru seems to be $t_{2g}^3$ with a high spin state, 
namely close to Ru$^{5+}$.
Generally, the superexchange interaction between fully occupied 
$t_{2g}^3$ states in nearest neighbor magnetic ions tends to be
stabilized with an antiferromagneticic ordering of their moments.
On the other hand, 
the double exchange mechanism through itinerant electrons
may not be favorable, 
since the majority $t_{2g}$ states are strongly localized and 
almost fully occupied - as seen in Fig.~\ref{fig4} (b) and (c).
Therefore,
the magnetism of Ru in RuSr$_2$GdCu$_2$O$_8$ 
is determined by the superexchange mechanism.

\subsection{Band structure and Fermi surface}
The calculated band structures for 
RuSr$_2$GdCu$_2$O$_8$ are shown along high symmetry directions in Fig.~\ref{fig5}.
The majority and minority spin states in AFM-II are degenerate 
due to the symmetry of 
the antiferromagnetic Ru and Gd moment alignments, as shown in Fig.~\ref{fig2}.
In the AFM-I structure, 
a small difference between majority and minority spin states
was observed due to the assumed ferromagnetic Gd moments, 
but its difference is negligibly small.
The band structure of AFM-I is practically the same as that 
observed in the AFM-II structure.
The bands around $E_F$ for all three structures
arise from the Ru-O$_{Ru}$-O$_{apical}$ bands (solid light circles)
and Cu-O$_{Cu}$ bands (solid dark circles).
In the AFM-I and AFM-II structures,
there are three bands crossing $E_F$ -
as indicated in Fig.~\ref{fig5} (c).
They arise from a Ru-O$_{Ru}$ band 
with antibonding $d_{xy} -  p_{y}$ ($d p \pi$) orbitals,
and two Cu-O$_{Cu}$ bands composed of
antibonding $d_{x^2-y^2} -p_{x}$ ($d p \sigma$) obitals
with even and odd symmetry
derived by a mirror operation in the Cu-O$_{Cu}$ bilayer.
The two Cu-O$_{Cu}$ bands are almost the same as those in the FM structure.
As expected from the structural similarity between 
RuSr$_2$GdCu$_2$O$_8$ and YBa$_2$Cu$_3$O$_7$,
their band structures arising from Cu-O$_{Cu}$ layers are close to each other,
showing a strong two-dimensionality.
The energy states are little changed 
upon going from $\Gamma$ to Z along the $z$ direction.
Also, the strong $d p \sigma$ hybridization leads to a wide bandwidth
of 10.1, 9.7 and 9.7~eV for FM, AFM-I and AFM-II, respectively.
These values are almost the same as observed
in the 2D $dp\sigma$ bands in YBa$_2$Cu$_3$O$_7$, 9.0~eV.\cite{87Mas,91Yu}
The slightly larger bandwidth may be
due to the smaller interatomic distance (1.93~\AA) of Cu-O$_{Cu}$ bonds
and the smaller buckling angle (5.7$^\circ$) in the CuO$_2$ plane
of RuSr$_2$GdCu$_2$O$_8$,
compared with those of YBa$_2$Cu$_3$O$_7$ ({1.95~\AA}  
and 7.8$^\circ$).

From a comparison of band structures between the FM and AFM-I structures,
we found that 
the antibonding $d p \sigma$ bands
are not influenced significantly by the magnetism of Ru 
except around the M point in the zone.
Around M,
where they possess small components of Cu $d_{3z^2-r^2}$ character,
the Cu $e_g$ states couple to the Ru $e_g$ states
through the O$_{apical}$ $p_z$ orbitals,
but do not couple to Ru $t_{2g}$ states.
Therefore, the quite small exchange splitting in the $d p \sigma$ states 
may be mediated through a more indirect coupling path with
O$_{apical}$-O$_{Ru}$.
Note that the Cu $t_{2g}$ states are coupled to the Ru $t_{2g}$ states 
through the O$_{apical}$ $p_{x(y)}$ orbitals,
but the Cu $t_{2g}$ states are almost fully occupied, 
so the hybridization channel at $E_F$ is depleted.

The calculated Fermi surfaces (FS) at $k_z = 0$ for the AFM-II structure 
are shown in Fig.~\ref{fig6} (a).
For convenience, a sketch of the FS is given in Fig.~\ref{fig6}~(b),
which makes for an easier comparison with the high $T_c$ results.
Two closed FS centered at the zone corner 
arising from the Cu-O$_{Cu}$ bands and
a closed FS from the Ru-O$_{Ru}$ band, are found. 
The inner and outer FS (1 and 2 in Fig.~\ref{fig6} (b)) correspond to 
antibonding $d p \sigma$ states with even and odd symmetry 
under $z$ reflection, respectively.
Note that the FS has no exchange splitting, 
due to the antiferromagnetic ordering of the Ru and Gd moments,
implies that the superconductivity need no longer be of the FFLO type
- in contrast to the FM case.\cite{99Pic}
A folding in $k$-space due to the antiferromagnetic ordering of Ru
makes intersections between the bands (a, b and c in Fig.~\ref{fig6} (b)),
showing the band-band interaction.
As observed in the undistorted FM,\cite{99Pic}
the FS clearly demonstrates 
similar nesting structures 
with those in the other high $T_c$ cuprate superconductors\cite{87Mas,91Yu};
these will give rise to singularities in the generalized susceptibility, 
and lead to possible anomalous behavior of the electronic properties.
Further, the FS (3 in Fig.~\ref{fig6} (b)), derived from Ru $d_{xy}$ states,
shows a pocket of electrons centered at $\Gamma$,
which may lead to the superconductivity in the Cu-O$_{Cu}$ layer 
with hole-doped character, as observed in experiments\cite{99Tal,99Ber}

In contrast,
the FS in the FM structure is complicated due to the Ru-O$_{Ru}$ bands.
The majority spin Ru $d_{xz(yz)}$ bands (4 in Fig.~\ref{fig5} (a))
become more dispersive and cross at $E_F$ compared to the AFM-I structure,
which makes an additional hole pocket centered at $\Gamma$,
leading to be a less hole-doped character in Cu-O$_{Cu}$ layers.
Hence,
the FM structure may show low $T_c$ or non-superconductivity.

\section{summary}
We have performed first-principles FLAPW calculations 
to investigate magnetic structures in superconducting RuSr$_2$GdCu$_2$O$_8$.
Contrary to the previously ferromagnetic RuSr$_2$GdCu$_2$O$_8$ structure
proposed from susceptibility experiments\cite{99Tal,99Ber} 
and first-principles calculations,\cite{99Pic} 
we found that antiferromagnetic ordering of the Ru moments is 
energetically favored over ferromagnetic ordering.
Our results are consistent 
with those of a recent neutron diffraction experiment,\cite{00Lyn} 
in which the antiferromagnetic ordering of Ru moments was predominantly 
observed. 
The magnetism arises from Ru $t_{2g}$ states
through the superexchange mechanism. 
The Ru $t_{2g}$ states
have a very small interaction with the Cu $e_g$ states
- which results in quite a small exchange splitting 
in the antibonding $dp\sigma$ states.
The Fermi surfaces derived from the antibonding $dp\sigma$ states have 
similar nesting structures 
with those in the 2D high $T_c$ cuprate superconductors, 
which may give a basis for superconductivity in this material.
Further experimental and theoretical investigations are necessary
to confirm this conclusion.
In any case, 
the coexistence of antiferromagnetism and superconductivity 
in RuSr$_2$GdCu$_2$O$_8$ is clearly allowed.

\section*{Acknowledgements}
Work supported by the U.~S. Department of Energy, 
Office of Science (at Northwestern University under Grant No.~DE-F602-88ER45372
and at Argonne National Laboratory under Contract No.~W-31-109-ENG-38)
and by a grant of computer time at NERSC.

\begin{table}
\caption{
Total energy difference, $\Delta E$ (in meV/cell), 
and calculated magnetic moments, $m$ (in $\mu_B$),
in the MT sphere of each atom for the FM, AFM-I and AFM-II structures of RuSr$_2$GdCu$_2$O$_8$. 
The symmetry of the AFM-I and AFM-II structures makes two inequivalent sublattice sites 
for Ru, O$_{apical}$ and Cu.
}
\begin{tabular}{clccc}
           &              &  FM   &  AFM-I & AFM-II \\ \hline
$\Delta E$ &              & 50.5  &  5.5           &  0.0           \\ \hline
           & Ru           & 1.59  &  1.57 (-1.53)  & 1.55 (-1.55)   \\
           & O$_{Ru}$     & 0.11  &  0.00          & 0.00           \\
$m$        & O$_{apical}$ & 0.13  &  0.11 (-0.10)  & 0.11 (-0.11)   \\
           & Cu           & 0.004 &  0.007 (0.015) & 0.004 (-0.004) \\
           & O$_{Cu}$     & 0.004 &  0.006         & 0.000          \\
\end{tabular}
\label{table1}
\end{table}

\begin{figure}
\caption{
(a) Crystal structure of RuSr$_2$GdCu$_2$O$_8$ with P4/mbm space group; 
(b) Top view of Ru-O$_{Ru}$ layer.
}
\label{fig1}
\end{figure}

\begin{figure}
\caption{
Magnetic ordering of Ru and Gd magnetic moments for (a) FM, (b) AFM-I and (c) AFM-II 
structures of RuSr$_2$GdCu$_2$O$_8$,
where the Cu, O$_{Ru}$, O$_{Cu}$ and O$_{apical}$ atoms are not given.
}
\label{fig2}
\end{figure}

\begin{figure}
\caption{
Total density of states (DOS) and partial DOS
of Ru $d$, O$_{Ru}$ $p$, O$_{apical}$ $p$, Cu $d$ and O$_{Cu}$ $p$ orbitals
for the FM, AFM-I and AFM-II structures of RuSr$_2$GdCu$_2$O$_8$.
Solid and dashed lines represent majority and minority spin states, respectively.
A vertical dotted line denotes the Fermi level.
}
\label{fig3}
\end{figure}

\begin{figure}
\caption{
Projected density of states (DOS) in real space for Ru $d$, O$_{Ru}$ $p$ 
and O$_{apical}$ $p$ orbitals 
for the FM, AFM-I and AFM-II structures of RuSr$_2$GdCu$_2$O$_8$. 
The $x$ -axis for the Ru $d$ (O$_{Ru}$ $p$) DOS 
are chosen as directions to neighboring O$_{Ru}$ (Ru) sites. 
}
\label{fig4}
\end{figure}

\begin{figure}
\caption{
Calculated band structure along high symmetry directions 
for the FM, AFM-I and AFM-II structures of RuSr$_2$GdCu$_2$O$_8$.
The bands originating mainly in the Cu-O$_{Cu}$ layers 
are represented with solid dark circles.
}
\label{fig5}
\end{figure}

\begin{figure}
\caption{
(a) Calculated Fermi surfaces (FS) of RuSr$_2$GdCu$_2$O$_8$ 
for the AFM-II structure at $k_z = 0$ and 
(b) its schematic illustration, 
which makes for an easier comparison with the high $T_c$ results.
}
\label{fig6}
\end{figure}

\end{document}